\documentclass[a4paper,onecolumn,preprint]{emulateapj}
\slugcomment{accepted for publication in ApJ, 17 December 2005}

\shorttitle{Accretion Modes in Collapsars}

\shortauthors{Lee \& Ramirez-Ruiz}

\begin{document}

\title{Accretion Modes in Collapsars - Prospects for GRB Production}

\author{William H. Lee} \affil{Instituto de Astronom\'{\i}a,
Universidad Nacional Aut\'{o}noma de M\'{e}xico, Apdo. Postal 70-264,
Cd. Universitaria, M\'{e}xico D.F. 04510:wlee@astroscu.unam.mx}

\and

\author{Enrico Ramirez-Ruiz\footnote{Chandra Fellow}}\affil{School of
Natural Sciences, Institute for Advanced Study, Princeton, NJ, 08540:
enrico@ias.edu}

\begin{abstract}
We explore low angular momentum accretion flows onto black holes
formed after the collapse of massive stellar cores. In particular, we
consider the state of the gas falling quasi-spherically onto
stellar-mass black holes in the hypercritical regime, where the
accretion rates are in the range $10^{-3}\lesssim \dot{M} \lesssim 0.5
M_{\sun}\;{\rm s}^{-1}$ and neutrinos dominate the cooling. Previous
studies have assumed that in order to have a black hole switch to a
luminous state, the condition $l\gg r_{\rm g}c$, where $l$ is the
specific orbital angular momentum of the infalling gas and $r_{\rm g}$
is the Schwarszchild radius, needs to be fulfilled. We argue that
flows in hyperaccreting, stellar mass disks around black holes are
likely to transition to a highly radiative state when their angular
momentum is just above the threshold for disk formation, $l \sim 2
r_{\rm g}c$. In a range $ r_{\rm g}c <l< 2r_{\rm g}c$, a {\it dwarf}
disk forms in which gas spirals rapidly into the black hole due to
general relativistic effects, without any help from horizontal viscous
stresses. For high rotation rates $l \geq 2 r_{\rm g}c$, the
luminosity is supplied by large, hot equatorial bubbles around the
black hole. The highest neutrino luminosities are obtained for $l
\approx 2 r_{g} c$, and this value of angular momentum also produces
the most energetic neutrinos, and thus also the highest energy
deposition rates. Given the range of $l$ explored in this work, we
argue that, as long as $l \geq 2r_{\rm g}c$, low angular momentum
cores may in fact be better suited for producing neutrino--driven
explosions following core collapse in supernovae and $\gamma$ ray
bursts.
\end{abstract}

\keywords{accretion, accretion disks --- dense matter ---
  hydrodynamics --- gamma rays: bursts --- supernovae: general}

\section{Introduction}\label{intro}

The collapse of massive cores in evolved stars is clearly one of the
most energetic events in astrophysics, producing observable
electromagnetic, neutrino, and, in all likelihood, gravitational
signatures. When it was discovered that the accretion shock launched
from the proto--neutron star after core bounce would stall under a
wide range of conditions, energy transfer from the proto--NS to the
outer regions through neutrinos was invoked as a possible mechanism to
re--energize the shock wave and explode the star
\citep{bw85}. One--dimensional (spherically symmetric) models
including the relevant microphysics and detailed neutrino processes
have consistently failed to produce successful supernova explosions
\citep{rj00,letal01,tbp03,bt03}. This has led to the exploration of
other possible channels and physical mechanisms to transfer the energy
to the shock and power the explosion.

The rotation and magnetic field of the progenitor and the newborn
compact object are important ingredients for a global understanding
the explosion mechanism \citep{lw70,wmw02,akiyama03,abm05,tqb05,wmd05}
and the subsequent evolution of the system, as is convection
\citep{jm96}. Recent studies have addressed rotation in two
\citep{fh00} and three dimensions\citep{fw04,janka05}, although a
definitive conclusion is still not available. Any of these effects may
enhance the effective neutrino luminosity in the inner regions and
thus power the explosion. Implications range from the success or
failure of the explosion itself, the possible production of a
classical $\gamma$--ray burst, the imparted kick to the newborn
neutron star or black hole, and the generation of a strong
gravitational wave signal.

Unfortunately, determining the precise manner in which a star is
rotating at core collapse is extremely complicated, depending upon
evolutionary details such as mass loss from the main-sequence star and
the configuration of the magnetic field, to name but two issues
\citep{spruit02,heger05} which lead to substantial core spin--down in
late evolutionary phases. Binary interactions may also clearly affect
the angular momentum of a star prior to core collapse, through
spin--up torques by tidal interactions.

In the context of GRBs, the collapsar model \citep{woosley93} invokes
the formation of a massive accretion disk around a new-born black hole
through fall-back accretion after core collapse. The energy release
associated with the huge accretion rates (of order 0.1 solar masses
per second) may be sufficient, if focused along the rotation axis of
the star, to produce a GRB. The angular momentum of the infalling gas
is an important parameter in this case, since it determines if a disk
will form (if there is no rotation only Bondi--like accretion will
ensue), as well as its dimensions. The accretion disk is likely to be
small enough that general relativity will play an important role in
its inner regions, allowing accretion even for finite values of
angular momentum, as will be discussed below.  A critical concern when
assessing the viability of progenitors for the production of GRBs has
thus been their rotation rate, as can best be determined from stellar
evolution considerations. We note here that \citet{wh05} and
\citet{yl05} have recently pointed out a new evolutionary channel for
very massive stars, in which mass and associated angular momentum
losses are greatly reduced by thorough mixing on the main sequence and
the subsequent avoidance of the giant phase. Previous work in
collapsar models and the associated neutrino--cooled accretion flows
\citep{mw99,pwf99,npk01} has generally considered relatively high
rotation rates and associated angular momentum values, typically with
$l\approx 6 r_{g} c$. With these values large, centrifugally supported
disks several hundred kilometers across promptly form around the
newborn black hole.\footnote{The production of magnetically driven
outflows has been considered as well, see \citet{proga03}} In this
paper we consider a wide range of rotation rates, covering cores with
near radial inflow to values slightly below those usually considered
in collapsar calculations. In the low-angular momentum limit, we find
that considerable energy release is still possible, and could
contribute significantly to the energy release available for a GRB.
Some elements of this picture have been considered in a quite
different context before, namely High--Mass X--ray Binaries, and we
draw some analogy and useful comparisons from these studies. In
\S~\ref{angmom} we detail the consequences of angular momentum on the
flow of collisionless matter, and note the regime where hydrodynamics
comes into play. A presentation of the different accretion modes that
occur, depending on whether angular momentum transport plays a role or
not is made in \S~\ref{acc}. We conclude in \S~\ref{disc} with a
discussion on the prospects for GRB production from such systems.

\section{The consequences of angular momentum}\label{angmom}

Before proceeding to the particular results of this set of
calculations, relevant for collapsars, this section is devoted to
more general considerations. The reader may wish to also consult the
clear introductory exposition by \citet{bi01}.

\subsection{Flow lines for ballistic parabolic motion}\label{ballistic}

Consider purely ballistic motion of collisionless matter with zero
energy in the gravitational potential well of a central mass, $M$. In
Newtonian theory, a test mass will follow a parabolic trajectory,
uniquely determined by its angular momentum, $l$. If the central mass
is a mathematical point, this will occur for any value of $l$. In a
realistic astrophysical situation, the mass $M$ has a finite radius
$R$, so capture orbits exist whenever the periastron distance is
smaller than $R$. One may term accretion in such a situation ``direct
accretion'', or accretion by capture. The effective cross section of
the mass $M$ for this capture, $\sigma_{\rm Newt}=\pi
R^{2}(1+2GM/v_{\infty}^{2}R)$, where $v_{\infty}$ is the particle's
velocity far from the mass M, is larger than the physical cross
section, $\sigma=\pi R^{2}$, because of gravitational focusing
\citep{st83}.

In general relativity (GR), there is an additional effect, because the
centrifugal barrier disappears even for finite $l$. Any particle with
enough energy at a given angular momentum will inevitably fall onto
the central mass. For definiteness, if $l=2r_{g}c$, the effective
potential exhibits a local maximum at $r=4GM/c^{2}$ (where an
unstable, marginally bound circular orbit is possible). A particle in
this situation with energy $E>0$ may thus directly accrete. Capture
orbits are thus of a different nature than in Newtonian theory, and
the effective cross section for this to occur may be written as
$\sigma_{\rm GR}=4\pi(2M)^{2}/v_{\infty}^{2}$, which corresponds to a
critical angular momentum $l_{\rm crit}=4GM/c$.

\begin{figure}
\plotone{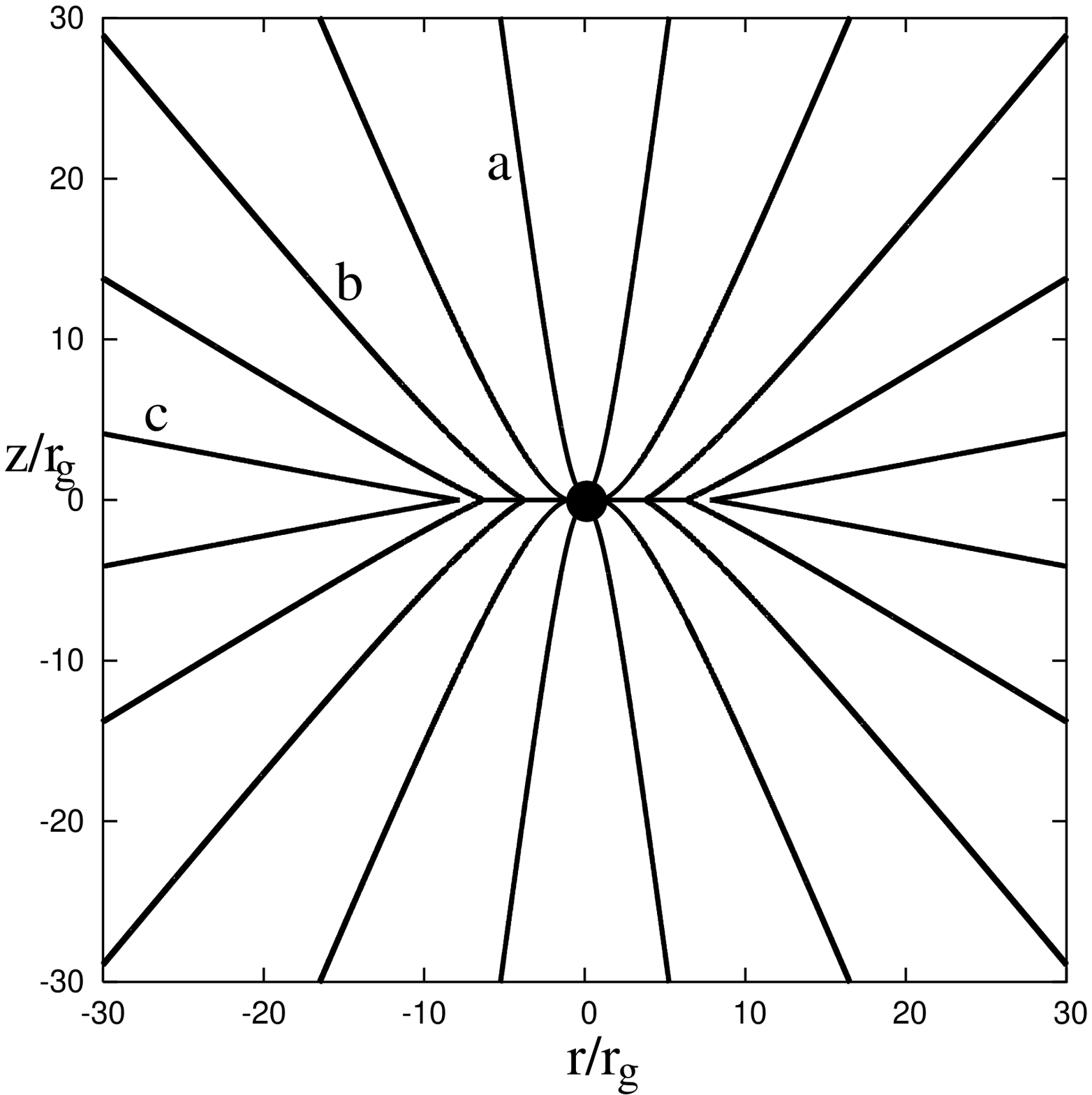}
\caption{Flow lines for ballistic, nearly parabolic motion of
particles surrounding a black hole of mass $M$ and rotating about a
given axis, $z$, with angular frequency $\Omega$. It is assumed that
the specific angular momentum, $l$, increases monotonically with the
polar angle $\theta$, and that flow lines do not intersect before
reaching the equator. The particular lines drawn here are for rigid
body rotation, with $l=l_{0} \sin^{2} \theta$. Lines of type (a)
directly impact the black hole, those of type (b) would do so as well
if they did not cross the equator first, while those of type (c) have
enough angular momentum to remain in equatorial orbit at a finite
radius $r$. If the energy in vertical motion is dissipated
efficiently, a thin inviscid disk will form in the equatorial plane,
denoted by a thick line.
\label{flowlines}}
\end{figure}

Consider now a situation in which a rotating cloud of particles, with
angular momentum about the z--axis, enshrouds a central black hole of
mass $M$. For vanishing particle energy, these will be in near
free-fall and follow approximately parabolic orbits. For a
monotonically increasing distribution of $l$ with the polar angle
$\theta$, particles closer to the poles will fall more easily than
those along the equator. We may divide the trajectories into three
categories, shown in Figure~\ref{flowlines}: (a) those that accrete
directly; (b) those that would accrete onto the black hole directly,
but cross the equatorial plane, $z=0$, before they are able to do so;
and (c) those that encounter their circularization radius (where the
centrifugal force balances gravity) on the equatorial plane and will
not accrete directly if the equatorial angular momentum is high
enough. Lines of type (b) coming from one hemisphere in fact encounter
an approaching line from the opposite direction. If the energy in
vertical motion is dissipated efficiently (through radiation) a thin
disk will form. The material in this disk does not have sufficient
angular momentum to remain in orbit, and will thus fall onto the
central mass in a short timescale, even in the absence of any
mechanism that transports angular momentum. This particular scenario,
with Compton cooling in mind, was applied by \citet{bi01} in the
context of high--mass X--ray binaries (HMXBs). That the flow lines
corresponding to ballistic motion do indeed cross obviously indicates
that a shock may form, depending on the local sound speed and fluid
velocity, and that a full analysis requires the inclusion of
hydrodynamical effects.

\subsection{Hydrodynamical effects}\label{hydro}

Initially, a shock may form in the equatorial plane, $z=0$, and the
energy in vertical motion will be transformed into internal energy. If
cooling occurs, some of this energy will be lost from the system
(through photons or neutrinos, depending on the physical
conditions). If the cooling rate $\dot{q}$ is able to balance the
energy input behind the shock front, the disk will remain
geometrically thin, with a scale height $H \ll r$. The fluid will then
orbit the central mass until it reaches a circular orbit corresponding
to its angular momentum. In Newtonian theory, this is the
circularization radius $r_{\rm c}=l^{2}/GM$. Thereafter, if angular
momentum is removed, it may slowly accrete onto the central mass
through a sequence of quasi--Keplerian orbits. In the case of a
relativistic potential, as considered here, only the orbits of type
(c) described in \S~\ref{ballistic} will be like this. Type (b)
trajectories will remain in the equatorial plane after passing through
the shock front, but the fluid along them will never find a proper
circularization radius, because of its low value of angular
momentum. This will form the fast, inviscid and near free--fall disk
envisaged by \cite{bi01}.

If the cooling mechanism is not so efficient, and is unable to
dissipate all the internal energy generated at the shock, a hot
toroidal bubble will form and grow in the equatorial plane. Its size
and stability depend on several factors. The first of these is simply
the total energy input, fixed by the accretion rate at the outer
boundary. The second is the cooling rate, given by the relevant
electromagnetic and neutrino processes (in the context of collapsars
only neutrino cooling plays a role). If the cooling rate does not drop
precipitously with decreasing density or temperature, the bubble may
eventually stabilize and reach a stationary configuration. A larger
bubble offers a larger volume within which to dissipate the internal
energy produced at the shock. The third is the geometry of injection
at the outer boundary, which directly affects the flow. For example,
if the injection is restricted to the equatorial plane
\citep{chen97,igu99}, polar outflows may appear, which transport
energy from the equatorial regions to higher latitudes. Finally, the
efficiency of angular momentum transport may cause enough advection of
matter and its accompanying internal energy so as to stabilize the
flow, or at least keep the bubble below a certain maximum size, even
if it is variable on short timescales. For the case of inviscid
hydrodynamic accretion, \citet{pb03a} have showed that the formation of
a hot torus directly affects the accretion rate onto the central
object. The fluid that would otherwise reach the equator at large
radii is deflected by the torus into a narrower funnel close to the
rotation axis.

Considering these further complexities, namely: (i) ``partial''
cooling, in which not all the internal energy generated at the shock
is removed from the system and (ii) a finite optical depth, which acts
in a similar way, it is clear that a full analytic solution cannot be
obtained, and that numerical calculations are necessary to study this
scenario.

\subsection{Input physics and initial conditions}\label{physics}

We believe the following assumptions will not alter the significance
of our overall results. We assume azimuthal symmetry and perform our
calculations in cylindrical $(r,z)$ coordinates. This allows for
greater spatial resolution than a 3D calculation and a solid
discussion of angular momentum effects. We do {\em not} assume
reflection symmetry with respect to the equatorial plane. We take the
fall back fluid to be in free-fall (i.e. on parabolic orbits) in the
central potential well of the black hole and superimpose on it an
angular momentum distribution corresponding to rigid--body rotation,
$l=l_{0} \sin^{2}\theta$, where $\theta$ is the polar angle, measured
from the rotation axis. $l_{0}$ is small compared to the local
Keplerian value needed to maintain a circular orbit, so it is nearly
in free--fall. Capture orbits are an essential ingredient in this
model, so we use the formula of \citet{pw80} for the gravitational
potential of the black hole, $\Phi(r)=-GM_{\rm BH}/(r-r_{g})$, where
$r_{\rm g}=2GM_{\rm BH}/c^{2}$ is the Schwarzschild radius.

This choice of matter and angular momentum distributions simplifies
the study of the problem, and its application to accretion following
core collapse, detailed below in \S~\ref{acc}. In a more detailed
treatment one could consider realistic stellar profiles in
pre--supernova cores (although these are derived in one dimension, and
mapping them to two or three dimensional dynamical studies is not a
trivial matter). The details of energy release, however, and the
driving of potential outflows are determined by processes in the
innermost, shocked regions of the flow, and are thus relatively
insensitive to the imposed external density profile. Furthermore,
during the relatively short timescales explored here, the variation of
these profiles may be assumed to remain constant with reasonable
accuracy.

We consider an equation of state with contributions from radiation,
non--degenerate relativistic electron--positron pairs and
non--degenerate $\alpha$ particles and free nucleons. For the latter,
we assume nuclear statistical equilibrium and calculate the mass
fraction of photodisintegrated nuclei as \citep{qw96}:
\begin{equation}
X_{\rm nuc}=22.4\left(\frac{\rho [{\rm g}~{\rm
cm}^{-3}]}{10^{10}}\right)^{-3/4} \left(\frac{T
[K])}{10^{10}}\right)^{9/8} \exp(-8.2 \times 10^{10}/T [K]).
\end{equation}
If this expression results in $X_{nuc} \ge 1$ we set $X_{nuc}=1$. The
pressure is given by:
\begin{equation}
P=\frac{11}{12}aT^{4}+\left(\frac{1+3X_{nuc}}{4}\right) \frac{\rho k
T}{m_{u}},
\end{equation}
where all the symbols have their usual meanings.  We have assumed for
simplicity that the electron fraction is constant throughout, with
$Y_{\rm e}=0.5$, i.e., there is one neutron for every proton. Clearly
for very high densities this is not the case, as neutronization occurs
and lowers $Y_{\rm e}$, reaching $Y_{\rm e} \approx 0.3$ for $\rho
\approx 10^{10}$g~cm$^{-3}$. This has two direct effects: (i) the
contribution to the pressure from electrons is altered, and (ii)
cooling through $e^{\pm}$ capture onto free nucleons is modified. In
the range of densities and temperatures explored here, however, both
of these effects are quite small and neglecting them does not alter
the essence of our results (see, however, \citet{lrrp05} for a
detailed dynamical calculation which does consider these effects in
the context of post--merger accretion disks).

The gas is opaque to electromagnetic radiation, and the main source of
cooling (other than advection into the black hole) is neutrino
emission. Given the high temperatures and the degree of
photodisintegration, the dominant terms arise from e$^{\pm}$
annihilation and e$^{\pm}$ capture by free nucleons. The corresponding
cooling rates (consistently with the equation of state) are then
\citep{km02,itoh89}
\begin{equation}
\dot{q}_{\rm cap}=X_{nuc} \, 9.2 \times 10^{33}
\left(\frac{T[K]}{10^{11}}\right)^{6} \frac{\rho [{\rm g}~{\rm
cm}^{-3}]}{10^{10}} \, {\rm erg}~{\rm cm}^{-3}~{\rm s}^{-1},
\end{equation}
and
\begin{equation}
\dot{q}_{\rm pair}=4.8 \times 10^{33}
\left(\frac{T[K]}{10^{11}}\right)^{9} \, {\rm erg}~{\rm cm}^{-3}~{\rm
s}^{-1}.
\end{equation}
Nucleon--nucleon bremsstrahlung neutrino emission is also included in
the code, but is insignificant compared with the other
two. Photodisintegration of $\alpha$ particles is also taken into
account in the energy equation \citep[see ][]{lrrp05}.  The densities
and temperatures do not rise enough for the gas to be optically thick
to neutrinos, i.e. $\tau_{\nu} \ll 1$ always (this is estimated by
considering coherent scattering off free nucleons and $\alpha$
particles, as well as absorption by free nucleons). Thus all the
energy in the form of neutrinos is lost immediately, and the cooling
is extremely efficient in this respect.  The hydrodynamics is followed
using a 2D Smooth Particle Hydrodynamics code \citep{lrr02,lrrp04},
which includes all the terms from the viscous stress tensor and uses
an $\alpha$--prescription for the magnitude of the viscosity,
$\nu=c_{\rm s}^{2} \alpha/\Omega_{k}$ ($c_{\rm s}$ is the local sound
speed and $\Omega_{\rm k}$ is the Keplerian angular velocity).

In reality, neutrinos, once emitted, are capable of transporting
energy in the flow and depositing it through the inverse of the
processes just described. At the low optical depths in fact
encountered in our calculations, this will alter the internal energy
only in the innermost regions of the flow, and by less than
approximately one per cent. This energy deposition may be quite
relevant in the production and driving of outflows, but its effect on
the overall accretion flow is small enough that we neglect it in a
first approximation.

\section{Accretion Morphologies and Luminosities}\label{acc}

We now address the particular case of accretion following core
collapse in massive stars within the context outlined in
\S~\ref{angmom}. The relevant parameters for a calculation are the
equatorial angular momentum, $l_{0}$, the viscosity $\alpha$, the
accretion rate at the outer boundary, $\dot{M}$, and the location of
the boundary itself, typically at $50r_{g}$ (although for the highest
values of angular momentum we explored it was placed at $80
r_{g}$). We have concentrated on variations in the $l_{0}-\alpha$
plane for various values of $\dot{M}$. For all the calculations
presented here, the central black hole is assumed to contain $M_{\rm
BH} = 4 M_{\sun}$. At a radius $r=r_{\rm in}$, the fluid is removed
from the calculation and its mass is added to that of the black
hole. Since we wish to resolve the accretion flow in the region where
the effects of General Relativity become important, this inner
boundary is necessarily very close to the Schwarszchild radius, and
typically $r_{\rm in}=1.5-2 r_{g}$. This is much less than the value
employed by \citet{mw99} ($r_{\rm in}=50$~km, or about $6 r_{g}$ for
$M_{\rm BH}\approx 3 M_{\sun}$), which allowed them to follow the
calculations for over 10 seconds, while we have computed the evolution
for a few tenths of one second. Given these timescales and the
characteristics of the late--time launching of winds from the
accretion disks in their calculations (speeds of order
20,000~km~s$^{-1}$ over distances of $\approx$~10,000~km), we do not
expect such outflows to occur in the present set of simulations.

\subsection{Low angular momentum -- Dwarf Disks}\label{dwarf}

If the angular momentum of the infalling gas is low, a substantial
fraction of the fluid is accreted directly onto the black hole, and
the rest produces an equatorial shock, where the energy in vertical
motion goes into thermal energy. It is efficiently radiated away in
neutrinos (recall that the densities are not high enough for neutrino
opacities to play any role), and a thin, dwarf disk forms. The actual
simulation looks very much like the analytical streamlines for
collisionless matter plotted in Figure~\ref{flowlines}. After an
initial transient lasting $\approx 200 r_{\rm g}/c \approx 6~$ms
(corresponding to the free--fall time of the material that is close to
the black hole), the system reaches a stationary state, in which as
much energy is dissipated in the shock as is radiated in neutrinos.

\begin{figure}
\plotone{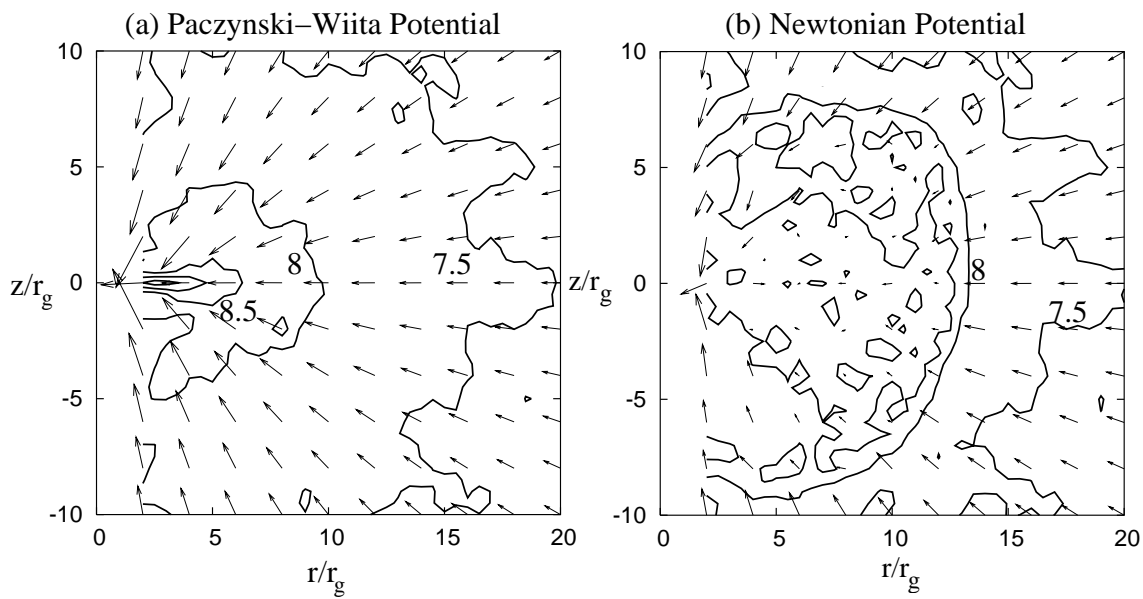}
\caption{Logarithmic density contours and velocity field in the
meridional plane for low angular momentum calculations ($l_{0}=1.9
r_{g} c$). The accretion rate is $\dot{M}=0.5~M_{\sun}$~s$^{-1}$. (a)
In the pseudo--Newtonian potential of Paczy\'{n}ski \& Wiita, the flow
is steady and the character of the solution is independent of the
adopted value of the viscosity parameter $\alpha$ and of the accretion
rate. Only a thin dwarf equatorial disk is present at small radii. (b)
In a Newtonian potential an accretion shock is formed as the gas
encounters the centrifugal barrier and a hot toroidal bubble is
created. The contours are evenly spaced every 0.5 dex and labeled in
units of g~cm$^{-3}$. }
\label{bubbleslowl}
\end{figure}

The steady, low--angular momentum flow configuration is insensitive to
the actual value of the viscosity parameter that is used. Calculations
with and without viscosity ($\alpha=0.1$ and $\alpha=0$) for
$l_{0}=1.9 r_{g} c$ result in an identical structure, shown in
Figure~\ref{bubbleslowl}a. The system is in the {\em inviscid} regime,
in which accretion onto the central mass is driven by general
relativistic dynamical effects, and not by the transport of angular
momentum. In the Newtonian regime this solution is non--existent
because of the lack of capture orbits for finite $l$.  Results of a
test calculation with the same initial conditions as just described,
but in a background Newtonian potential, where $\Phi(r)=\Phi_{\rm
N}(r)=-GM_{\rm BH}/r$ are displayed in Figure~\ref{bubbleslowl}b. As
expected, the infalling gas is perfectly able to settle at the
circularization radius and the dwarf disk never forms. Instead a
shocked, hot toroidal bubble is created, akin to those found in
high--angular momentum calculations described below in
\S~\ref{bubbles}.

The neutrino luminosity as a function of time is fairly constant,
although fast variations at small amplitude are apparent. Part of this
variability is of numerical origin, because the equatorial dwarf disk
is so thin that it is difficult to resolve adequately (the noise level
decreases if the resolution of the calculation is increased). The
density in this type of flow is essentially given by the free--fall
type conditions and mass conservation through the continuity equation,
as
\begin{equation}
\rho \approx \frac{\dot{M}}{4 \pi r^{2} v_{r}} = \frac{\dot{M}}{4 \pi
r_{g}^{2} c} \left( \frac{r}{r_{g}} \right)^{-3/2} = 2 \times 10^{9}
\left( \frac{\dot{M}}{0.5 M_{\sun} s^{-1}}\right) \left(
\frac{r}{r_{g}} \right)^{-3/2} {\rm g}\, {\rm cm}^{-3}
\end{equation} 
for $M_{\rm BH}=4 M_{\sun}$. The crucial difference from spherical
Bondi--type accretion is in the formation of the equatorial shock
because of the finite, albeit small, value of angular momentum, which
raises the temperature to $\simeq 3-4$~MeV. Thermal effects dominate,
and the mean neutrino energy is accordingly of order $E_{\nu} \approx
4 kT$.

\subsection{High angular momentum -- Toroidal Bubbles}\label{bubbles}

\begin{figure}
\plotone{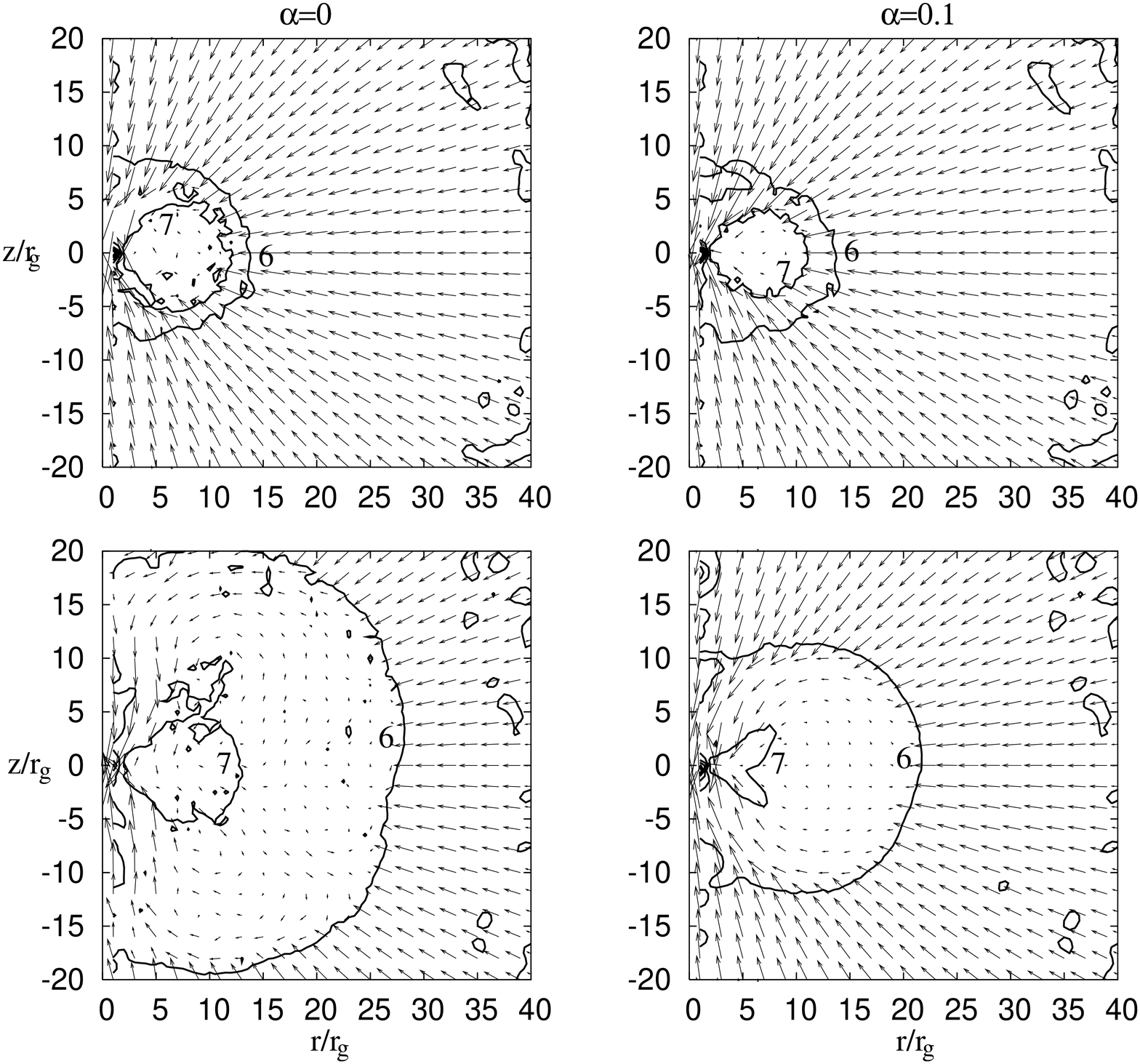}
\caption{Logarithmic density contours and velocity field in the
meridional plane for high angular momentum calculations ($l_{0}=2.1
r_{g} c$), with $\alpha=0$ (left column) and $\alpha=0.1$ (right
column), at $t=500 r_{g}/c$ (top row) and $t=1500 r_{g}/c$ (bottom
row). The accretion rate is $\dot{M}=0.01~M_{\sun}$~s$^{-1}$. In both
cases the hot toroidal bubble grows continously as the shock front
moves outward, but it is clearly smaller in the high--viscosity
case. The contours are evenly spaced every dex and labeled in units of
g~cm$^{-3}$.}
\label{bubbleshighl}
\end{figure}

\begin{figure}
\plotone{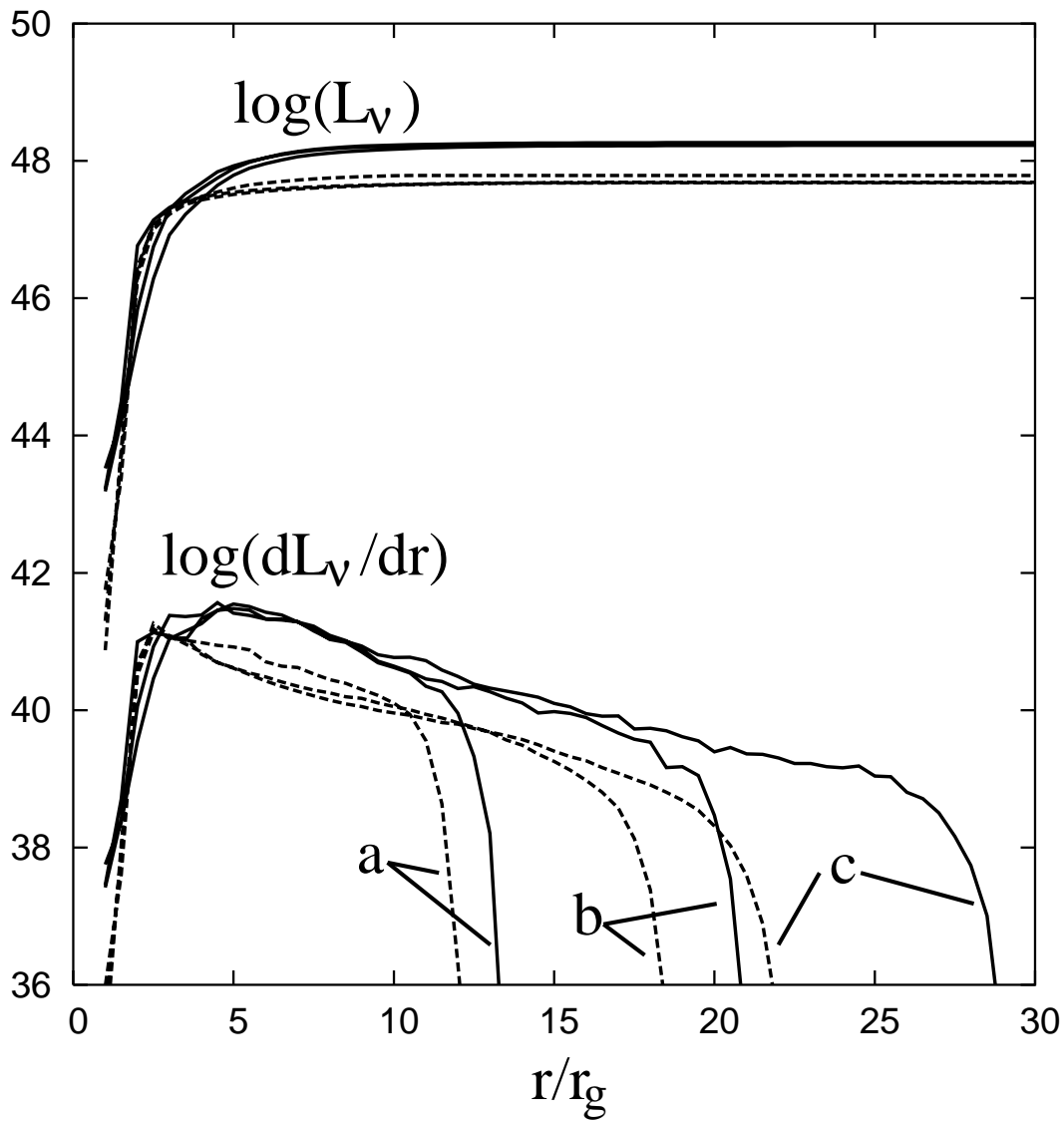}
\caption{Neutrino luminosity per radial interval, $dL_{\nu}/dr$ (lower
lines), in units of erg~cm$^{-1}$~s$^{-1}$, as a function of radius at
(a) $t=500 r_{g}/c$, (b) $t=1000 r_{g}/c$, and (c) $t=1500 r_{g}/c$
for high--angular momentum calculations ($l=2.1 r_{g} c$) without
($\alpha=0$, solid lines) and with ($\alpha=0.1$, dashed lines)
viscosity, for $\dot{M}=0.01$~M$_{\sun}$~s$^{-1}$. The corresponding
solid and dashed lines above show the integrated luminosity $L_{\nu}$
in erg~s$^{-1}$. The outward motion of the shock is clearly visible in
the sharp drop in emissivity as the calculation progresses. }
\label{Lnuvsr}
\end{figure}

For high values of $l_{0}$, the material closest to the equator has
enough angular momentum to remain in stable circular orbit around the
black hole. We first consider $\dot{M}=0.01$M$_{\sun}$~s$^{-1}$, and
defer higher accretion rates to \S~\ref{mdot}. A shock initially forms
close to the circularization radius, and the energy in free--fall is
transformed into internal energy. The temperature and density in the
gas rise rapidly behind the shock front, to a few~$\times 10^{10}$~K,
and $\approx 10^{7}-10^{9}$~g~cm$^{-3}$, depending on the assumed
value of $\dot{M}$. A hot toroidal bubble is promptly formed and
intense neutrino emission takes place within it (see
Figure~\ref{bubbleshighl}). The material is rapidly photodisintegrated
into nucleons and protons after passing through the shock, and
$e^{\pm}$ annihilation and capture onto free nucleons contribute to
the total neutrino luminosity. The global behavior of the accretion
flow depends essentially on the efficiency of angular momentum
transport, and also on the accretion rate. For high viscosities
($\alpha=0.1$), the gas is accreted efficiently so that the growth of
the bubble is slower, and the total neutrino luminosity is {\em lower}
by about a factor of 2 with respect to the inviscid ($\alpha=0$)
case. This comes about because: (i) a large fraction of the bubble is
hot enough to radiate copious amounts of neutrinos, and high viscosity
implies a smaller radiating volume due to the efficient transport of
angular momentum (see Figure~\ref{Lnuvsr}); and (ii) more efficient
angular momentum transport means that less matter accumulates in the
equatorial region inside the bubble, and thus less energy is released
in neutrinos. Note that the fact that a higher viscosity also implies
a greater amount of dissipated energy does not reverse this trend.

\begin{figure} 
\plotone{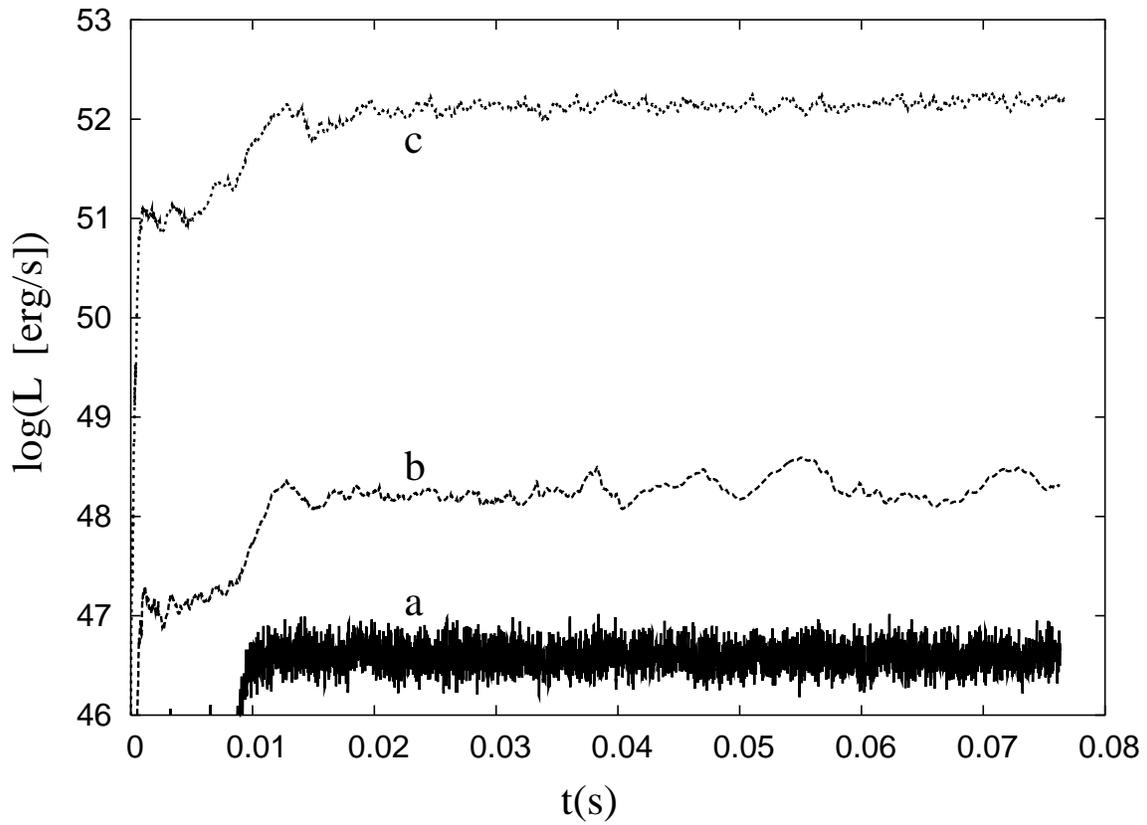}
\caption{Neutrino luminosity $L_{\nu}$ in erg~s$^{-1}$ as a function
of time for inviscid ($\alpha=0$) runs with (a) $l_{0}=1.9 r_{g} c$,
$\dot{M}=0.01$~M$_{\sun}$~s$^{-1}$; (b) $l_{0}=2.1 r_{g} c$,
$\dot{M}=0.01$~M$_{\sun}$~s$^{-1}$; (c) $l_{0}=2.1 r_{g} c$,
$\dot{M}=0.5$~M$_{\sun}$~s$^{-1}$. For low angular momentum it is
essentially steady, although strong variability around a steady
average is apparent. For high angular momentum, high amplitude,
quasi-periodic variations are seen as a result of large--scale
variations in the size and shape of the hot toroidal bubble.}
\label{Lnuvst}
\end{figure}

The neutrino luminosity for $l_{0}=2.1 r_{g} c$ and
$\dot{M}=0.01$M$_{\sun}$~s$^{-1}$ as a function of time is shown in
Figure~\ref{Lnuvst}. There are large--amplitude fluctuations, due to
variations in the size, shape and structure of the growing toroidal
bubble. Changes by up to a factor of two occur over a background
emission rate that is fairly steady, and on a timescale $\Delta t
\approx r_{\rm b}/c_{\rm s}$, where $r_{\rm b} \approx 20 r_{\rm g}$
is the size of the hot bubble, and $c_{\rm s}$ is the sound speed. The
temperature is typically $T \approx 10^{10}$~K, so this gives $\Delta
t \approx 2$~ms, in rough agreement with the observed variability
timescale of $\approx 150 r_{\rm g}/c=4-5$~ms. Most of the emission
arises from the innermost equatorial regions, and it is changes in
this volume that affect the total luminosity.

In the inviscid limit there is no angular momentum transport, and so a
substantial fraction of the matter cannot accrete. The toroidal bubble
thus grows, fluctuating on short timescales and eventually becoming as
large as the computational domain (at which point the calculation is
stopped). We note that outflows are ubiquitous in this case, as they
are a natural way of transporting the stored energy to larger radii. A
detailed analysis of their characteristics is beyond the scope of this
paper and is left for future work. If the viscosity is finite,
however, the bubble eventually reaches a steady state, in which the
accretion by angular momentum transport is balanced by the mass and
energy inflow at the outer boundary.

For a given magnitude of the viscosity, the size of the bubble depends
on the equatorial angular momentum, $l_{0}$. The circularization
radius $R_{\rm c}$ for matter falling along $z=0$ in the
pseudo--Newtonian potential is determined by the condition $R_{\rm
c}^{3}/[R_{\rm c}-r_{g}]^{2}=l_{0}^{2}/GM_{\rm BH}$, and gives a first
estimate of the shock location. All other things being equal, at large
angular momentum the torus will be larger, and of lower density and
temperature. Despite the larger neutrino--emitting volume and due to
the steep dependence of the cooling rates on the temperature, for a
fixed magnitude of the viscosity, larger tori have lower luminosities.

\subsection{Varying the accretion rate}\label{mdot}

We have considered variations in the accretion rate by computing
models with $\dot{M}=10^{-3}, 10^{-2}, 10^{-1}, 5 \times
10^{-1}$~M$_{\sun}$~s$^{-1}$. An increase by one order of magnitude
affects the density by about the same amount, simply because of mass
conservation. The temperature also rises, but only about 10$^{1/4}$
per order of magnitude increase in density. Since the neutrino
emission rates from pair capture and annihilation scale as $T^{9}$ and
$\rho T^{6}$ respectively, the total luminosities should increase by
$\approx 10^{2.5}$ and $10^{2.25}$. So the total change is by a
factor $\approx 10^{2.5}$. Indeed this is what is seen when the actual
values for $L_{\nu}$ are computed, so that at a fixed $l_{0}$,
$L_{\nu} \propto \dot{M}^{2.5}$. At very high accretion rates the
pre--shock region is dense and cold enough that degeneracy effects
start to become important. The post--shock region, however, is always
hot enough that our assumptions concerning the equation of state are
still valid.

\begin{figure}
\epsscale{0.6}
\plotone{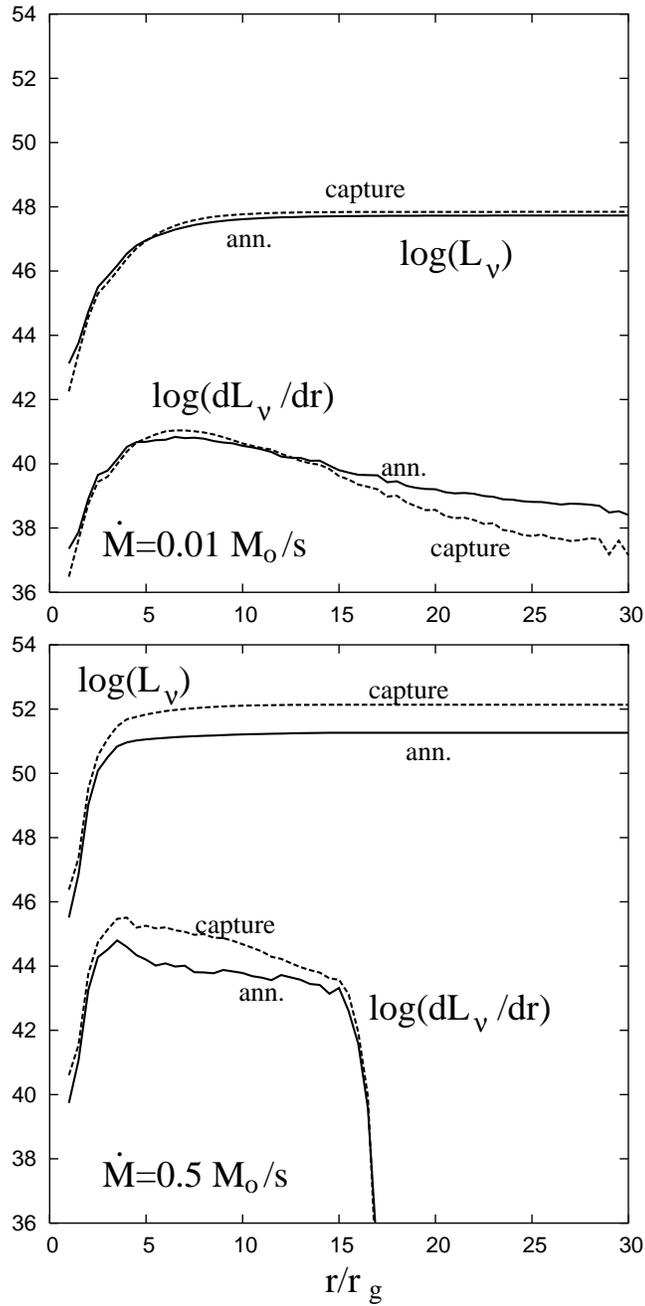}
\caption{Same as Figure~\ref{Lnuvsr}, but for $l=2.2 r_{g} c$ and
$\alpha=0$.  Calculations with $\dot{M}=0.01$~M$_{\sun}$~s$^{-1}$
(top) and $\dot{M}=0.5$~M$_{\sun}$~s$^{-1}$ (bottom) are shown at
$t=1500 r_{g} c$. The solid (dashed) lines show the contribution from
pair annihilation (pair capture onto free nucleons) to the total
luminosity. At high accretion rates pair capture dominates the cooling
and makes the hot toroidal bubble smaller.}
\label{Lnuvsrinviscid}
\end{figure}

\begin{figure}
\plotone{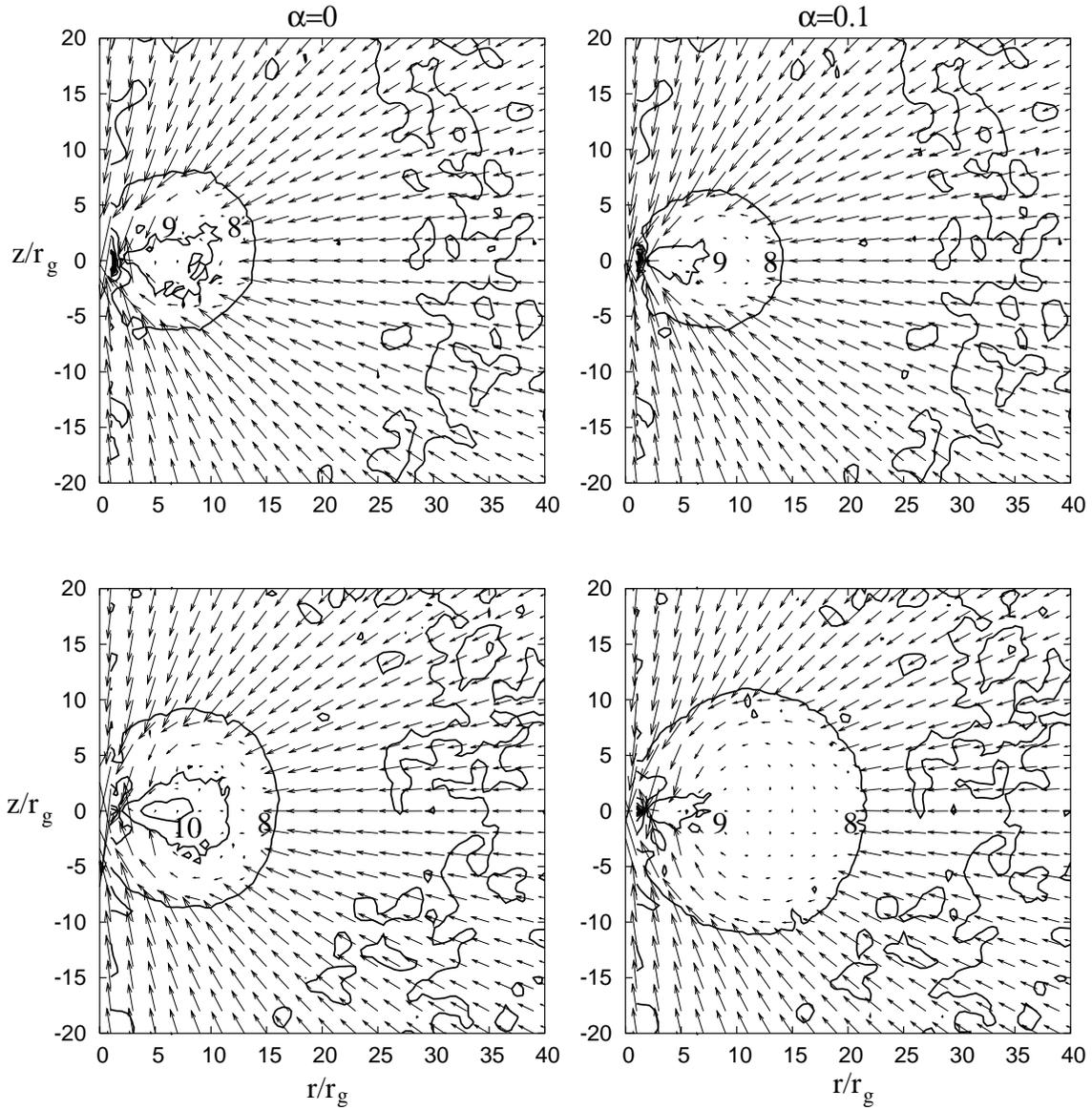}
\caption{Same as Figure~\ref{bubbleshighl} but for
$\dot{M}=0.5~M_{\sun}$~s$^{-1}$ and $l_{0}=2.1 r_{g} c$. As before,
the hot toroidal bubble grows as the shock front moves outward, but
this time, as opposed to the case of low accretion rate, it is clearly
smaller in the inviscid case. The contours are evenly spaced every dex
and labeled in units of g~cm$^{-3}$.}
\label{bubbleshighlhighmdot}
\end{figure}

For the highest accretion rates, $\dot{M}\simeq 0.1-0.5$
~M$_{\sun}$~s$^{-1}$, likely to occur in the initial stages of
fallback following the formation of the black hole, the solution
changes quantitatively from that described above in \S~\ref{bubbles},
because of the increased density. The annihilation of $e^{\pm}$ is
essentially a thermal process, and as such its emissivity depends only
on the temperature, while pair capture onto free nucleons involves
interactions with the background fluid, making the emissivity
proportional to the density. As the density increases, pair capture
becomes increasingly important until it eventually dominates the total
cooling rate by a factor of four at
$\dot{M}=0.5$~M$_{\sun}$~s$^{-1}$. This is illustrated in
Figure~\ref{Lnuvsrinviscid}, where the differential and integrated
neutrino luminosities are shown for two different accretion rates at
the same equatorial angular momentum (separated into the two main
processes contributing to the cooling rate). The net result is that
calculations without efficient transport of angular momentum reach
higher densities and emissivities, and thus produce smaller equatorial
bubbles than if viscosity is present (see
Figure~\ref{bubbleshighlhighmdot}).

\begin{figure}
\plotone{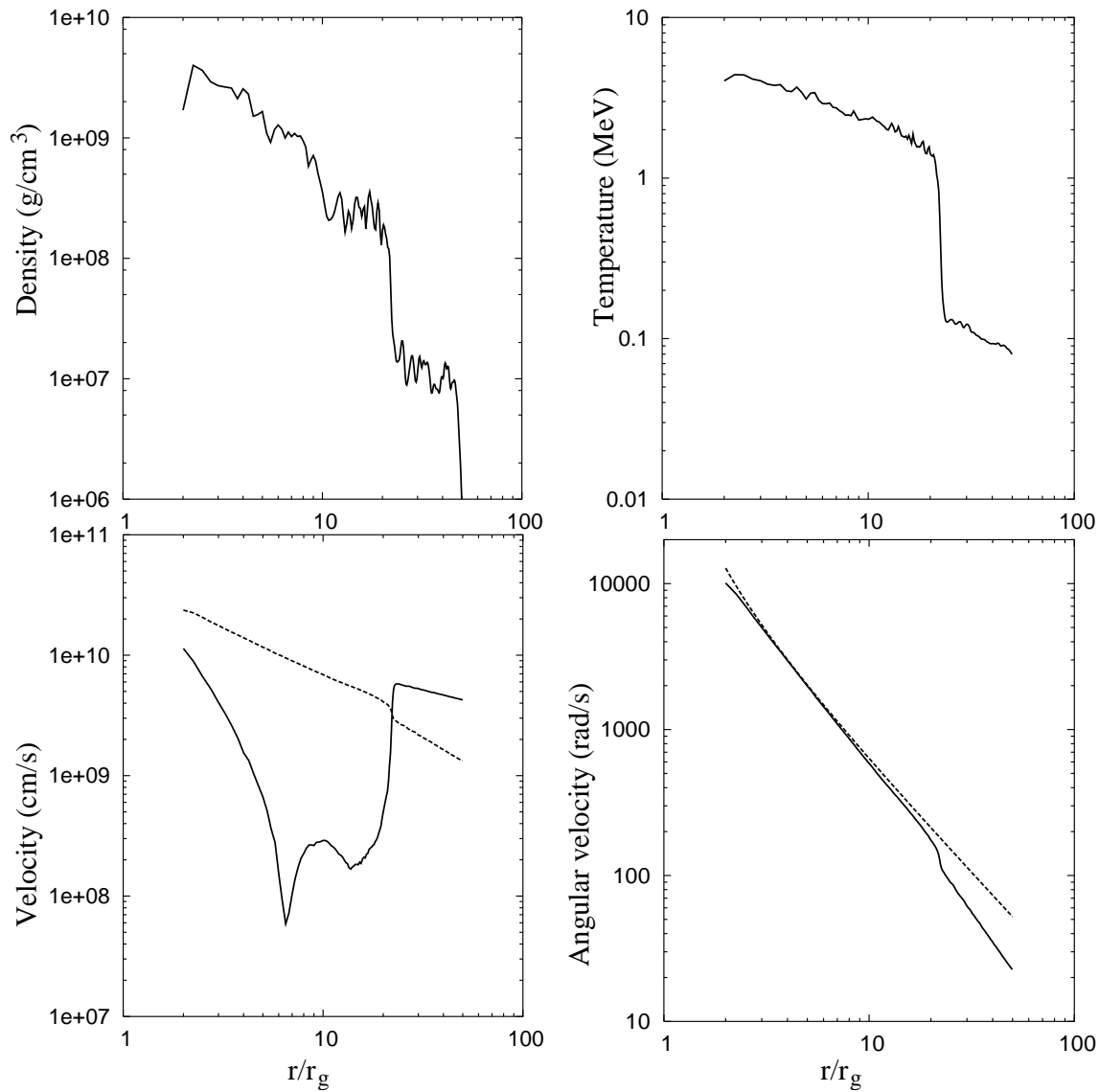}
\caption{Density, temperature and velocities along the equator, $z=0$,
for the calculation with $l_{0}=2.2 r_{g}c$, $\alpha=0.1$ and
$\dot{M}=0.5~M_{\sun}$~s$^{-1}$ at late times, when the toroidal
bubble has stabilized. In the bottom left panel, the solid (dashed)
line is for the radial (azimuthal) component of the velocity. The
bottom right panel shows the computed angular velocity (solid line) as
well as the Keplerian solution in the Paczy\'{n}ski--Wiita potential
for reference (dashed line).}
\label{profiles}
\end{figure}

As for lower accretion rates, with finite viscosity the hot torus
eventually stabilizes, with a total volume determined by the
equatorial angular momentum. Figure~\ref{profiles} shows the radial
run of several variables along $z=0$ for such a stable
configuration. The accretion shock is clearly seen as a large jump in
density and temperature. Within the torus, the rotation curve is
nearly Keplerian, and the radial velocity shows variations due to the
stable, large scale circulation pattern that is established and
maintained (see Figures~\ref{bubbleshighl} and
\ref{bubbleshighlhighmdot}). These solutions are similar to those
obtained analytically by \citet{pwf99} and numerically by \citet{mw99}
and \citet{proga03}. In the region exterior to the shock front they
differ because we have assumed pure free fall, whereas their models
consider either an extended Keplerian disk or an actual infalling
stellar envelope with substantial pressure support. This effective
boundary condition has little effect on the structure and evolution of
the inner accretion flow.

\begin{figure} 
\plotone{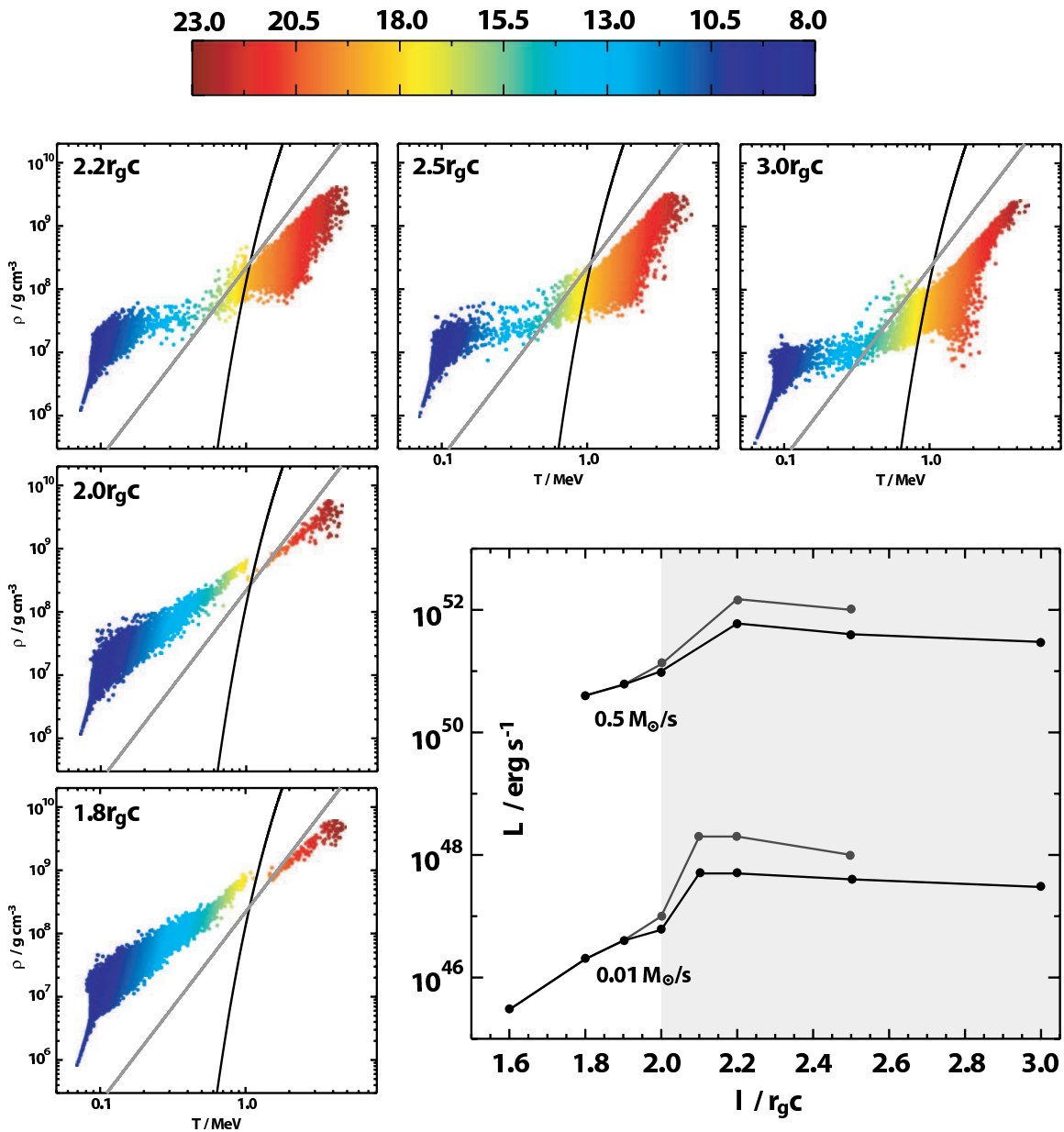} 
\caption{Flow structure in the density--temperature plane for
calculations with varying equatorial angular momentum $l_{0}$
(labeled) at $\dot{M}=0.5$~M$_{\sun}$~s$^{-1}$ and color coded
according to the volume cooling rate (in units of
erg~cm$^{-3}$~s$^{-1}$). The black and grey solid lines across each
plot show the degeneracy ($kT=7.7 \rho_{11}^{1/3}$~MeV) and
photodisintegration (50\% of $\alpha$ particles into free nucleons by
mass) thresholds respectively.  The transition from a dwarf disk to a
hot toroidal bubble as $l_{0}$ increases beyond $\simeq 2.1 r_{g} c$
is clearly seen, as is the slight decrease in maximum density at high
angular momentum. The characteristic neutrino luminosity is plotted in
the large panel as a function of $l_{0}$ for two different mass
accretion rates, and viscous ($\alpha=0.1$, black lines) and inviscid
calculations ($\alpha=0$, gray lines). The qualitative change in the
dependence of $L_{\nu}$ on $l_{0}$ occurs around $2.1r_{g}c$. Note
that at high accretion rates, even low angular momentum configurations
($l_{0}\simeq 1.8 r_{g} c$) are capable of releasing up to
$10^{51}$erg~s$^{-1}$.}
\label{rhoT} 
\end{figure}

Further insight can be gained about the structure of the accretion
flow by plotting the density as a function of the temperature (see
Figure~\ref{rhoT}). The location of the shock is clearly seen as a
rapid rise in temperature and density, separating two regions in the
$\rho-T$ plane. It is evident as well that most of the high--density
material within the hot toroidal bubbles at high angular momentum,
$l_{0}\geq 2.1 r_{g} c$, and the entire fluid in the dwarf disks in
the opposite limit, is entirely photodisintegrated (the corresponding
threshold is indicated by the black continous line in each
$\rho-T$ frame of Figure~\ref{rhoT}).

The physical conditions in the dwarf disks are largely insensitive to
the adopted value of $l_{0}$ at a fixed accretion rate, and we
typically find $\rho \approx 5 \times 10^{9}$~g~cm$^{-3}$ and $T
\approx 3-4$~MeV for $\dot{M}=0.5$ ~M$_{\sun}$~s$^{-1}$. The
total amount of mass under these conditions is, however, affected by
$l_{0}$, and impacts upon the absolute neutrino luminosity $L_{\nu}$.

The hot toroidal bubbles are evident in the panels of
Figure~\ref{rhoT} in the larger area of the $\rho-T$ plane that is
occuppied by the fluid, since there is a considerable gradient in both
temperature and density as one moves from the immediate post--shock
region to the inner equatorial disk. Note how the maximum density
slowly decreases as $l_{0}$ rises (at roughly constant temperature) as
the bulk of the equatorial inflow has a larger circularization
radius. This also affects the neutrino luminosity, although less
dramatically than in the case of dwarf disks for a comparable
fractional change in $l_{0}$.

Regardless of the accretion rate, when the system enters the {\em
inviscid} regime in the low angular momentum limit, $L_{\nu}$ becomes
independent of $\alpha$ and the luminosity drops rapidly. This can be
seen in the large panel in Figure~\ref{rhoT} where we plot the
luminosity as a function of the equatorial angular momentum $l_{0}$
for inviscid and viscous calculations at high and low accretion
rates. The joining of the two classes of solutions at $l_{0} \approx 2
r_{g} c$ marks the transition to inviscid accretion and the appearance
of the dwarf disk in near free--fall. For $l_{0} < 2 r_{g} c$,
$L_{\nu} \propto l_{0}^{16}$, reflecting the fact that a large
fraction of the infalling material is directly accreted by the black
hole, and thus the rapid shutoff as the angular momentum is
decreased. In the opposite case we find $L_{\nu} \propto l_{0}^{-2}$,
because larger toroidal bubbles are less dense and have lower
temperatures than smaller ones. Combining all of these results, we
find that the neutrino luminosity can be fitted piecewise as:

\begin{equation}
L_{\nu} \approx \left \{ 
\begin{array}{llll}
8 \times 10^{50} & \left( \frac{l_{0}}{1.9 \,
r_{g} c} \right)^{16} \left( \frac{\dot{M}}{0.5\, {\rm M}_{\sun} \,
{\rm s}^{-1}} \right)^{2.5} & \, {\rm erg}\, {\rm s}^{-1} & 
{\rm for } \, \, l_{0} \le 2 r_{g} c \\
& \\
10^{52} & \left( \frac{l_{0}}{2.2 \,
r_{g} c} \right)^{-2} \left( \frac{\dot{M}}{0.5\, {\rm M}_{\sun} \,
{\rm s}^{-1}} \right)^{2.5} & \, {\rm erg}\, {\rm s}^{-1} & 
{\rm for } \, \, l_{0} \ge 2 r_{g} c. 
\end{array}
\right .
\end{equation}

\subsection{Neutrino spectrum and energy deposition}\label{nuspectrum}

The shifting contributions to the cooling rate from different
processes as the accretion rate increases will affect the emergent
neutrino spectrum. Neutrinos produced by pair annihilation have
energies $E_{\nu} \approx 4 kT$. If degeneracy effects are important,
those arising from pair capture will have energies comparable with the
electron Fermi energy, and one can show that $E_{\nu} \approx 9
(\rho_{10} Y_{\rm e})^{1/3}$~MeV. For the models we have computed with
the highest accretion rates this is indeed the case. With these two
expressions we have computed the neutrino energies for models with
$\dot{M}=0.5$~M$_{\sun}$~s$^{-1}$ (see Table~1) as a function of the
equatorial angular momentum, $l_{0}$.  Altering $l_{0}$ clearly has an
effect on the energies of the emitted neutrinos.

For GRBs, the relevant deposition process is neutrino annihilation
$\nu_e+{\bar\nu_e} \rightarrow e^- + e^+$ with a corresponding
deposition rate given by
$Q^+_{\nu,{\bar\nu}}={\bar\sigma_0}\frac{L^2_{\nu}}{A^2} \left<
E_{\nu}\right>\,\zeta$ where ${\bar\sigma_0}=3K\,G^2_F/4$ is the
characteristic weak interaction cross section per unit energy squared,
$G^2_F=5.29\times 10^{-44}\,{\rm cm^2}\,{\rm MeV^{-2}}$ is the Fermi
constant, $K$ is a phenomenological electro--weak parameter usually
taken to be $0.1- 0.2$, $A$ is the surface area of the absorbing
region, and the multiplicative factor $\zeta$ takes into account the
geometry of the emitting region \citep{rrs05}. For a fixed amount of
energy release i.e., $L_{\nu}$, the efficiency for producing an
explosion increases with $\left< E_{\nu}\right>$.

\subsection{Accompanying winds, supernovae and $\gamma$--ray bursts}

As mentioned above (\S~\ref{physics}), we have not included the
effects of neutrino heating in our calculations, and believe that it
will not affect the behavior of the inner accretion flow. The
neutrinos may, however, ablate baryonic material from the surface of
the disk. The properties of such winds, arising from proto--neutron
stars following core--collapse, have been investigated before in
spherical symmetry \citep{qw96}. This is a different environment, so
we can consider it only as a very rough approximation for our
case. With the typical numbers from our simulations, we find that the
mass outflow rate will be approximately given by
\begin{equation} 
\dot{M}_b \sim 5 \times 10^{-4}\left({L_\nu \over 10^{52}\;{\rm
erg\;s^{-1}}}\right)^{5/3} M_\sun \: {\rm s^{-1}}.
\label{ablation}
\end{equation}
Most of the ablation is due to the absorption on nucleons. These
outflows are relatively slow. Even the gas that is ejected from the
innermost region of the disk has a speed of only $\sim0.1$c; the gas
that comes out from larger radii is even slower. This simple mass loss
rate estimate is comparable to that given by \citet{pth04}, where the
full hydrodynamic wind solution is given, assuming the analytic disk
structure essentially given by \citet{pwf99}. At $L_\nu \sim 3 \times
10^{52}\;{\rm erg\;s^{-1}}$, the ablation rate in
equation~(\ref{ablation}) is up to $\sim 10$ times higher than what
can be driven, on energetic grounds, by $\nu\bar{\nu}$ annihilation
alone. To see this, combine the lower bound to the photon-pair flux
estimated by the reaction $\nu + \bar{\nu} \rightarrow e^{+} + e^{-}$,
which yields $L_{\nu \bar{\nu}}\sim 10^{-3}L_\nu$ \citep{rrd03}, with
the estimate $\dot{M}_b \sim L_{\nu \bar{\nu}}(GM_*/R_*)^{-1} \sim 5
L_{\nu \bar{\nu}}/c^2$.

Powerful outflows can also be driven from accretion disks through
magnetic fields. If the magnetic fields do not thread the BH, then a
Poynting outflow can at most carry the gravitational binding energy of
the torus.  In this case, we may estimate their power as \citep{bp82}:
\begin{equation} L_{\rm MHD}=\frac{B^{2} r^{3} \Omega}{2} \approx
10^{52} \, {\rm erg} \, {\rm s}^{-1}, \end{equation} where we have
assumed that the field energy is at 10\% of equipartition with the
internal energy, $\rho c_{s}^{2}$, and typical values at the
marginally stable orbit for calculations with $\dot{M}=0.5
M_{\sun}$~s$^{-1}$ and $l_{0} \approx 2 r_{g} c$ ($\rho \approx
10^{10}$g~cm$^{-3}$, $kT \approx 5$~MeV). However, if the more massive
central black hole is rapidly spinning, its larger energy reservoir is
in principle extractable through MHD coupling to the disk by the
\citet{bz77} mechanism. These are clearly very rough estimates as none
of these effects is explicitly included in our calculations, but it is
clear that the large densities obtained even with modest values of the
equatorial angular momentum allow in principle substantial magnetic
fields to be anchored in the flow. Recent numerical MHD calculations
assuming both adiabatic and neutrino--cooled flows in collapsing
envelopes do show these general features
\citep{pb03b,mizuno04,proga05,mckinney05,dso05}.

Strong winds from hypercritical accretion flows may play an important
role in the production and morphology of supernovae and $\gamma$--ray
bursts. In the first place, the corresponding energy, however
transferred to the outer layers, may actually explode the star and
power the supernova explosion itself \citep{mw99,m03,tqb05}. Second,
the toroidal geometry inherent in flows endowed with angular momentum
will break the spherical symmetry, and allow for higher efficiency
than purely radial inflow in terms of liberating the gravitational
energy in the system. Asymmetries may be reflected in the explosion
itself and its remnant or in the beaming of a possible $\gamma$--ray
burst, and can be quantified through line profiles \citep{mazzali05}
and collimation of the outflow \citep{pk01}. Third, if the accretion
flow cools inefficiently because it becomes so dense that the internal
energy is simply advected into the black hole
\citep{lrrp04,jpdc04,lrrp05}, it is possible that strong winds may
actually revive the stalled accretion shock and make it reach the
stellar envelope \citep{knp05}. The argument is analogous to that
invoked for classical Advection Dominated Accretion Flows, or ADAFs
\citep{ny94}, except that the processes involve neutrino, instead of
radiative cooling. Finally, the nucleosynthesis in such winds has been
the focus of a number of recent studies, particularly in what concerns
the production of $r$--process elements \citep{pth04}. 

If any strong winds or outflows, by whatever mechanism, are eventually
driven from the vicinity of the black hole or the surface of the
accretion disk, these can affect the incoming flow itself, by removing
both mass and angular momentum. While such feedback almost certainly
occurs, and could under certain circumstances play an important role
in the overall dynamics, it is outside the scope of the present paper
and is left as future work.

The observable consequences are thus potentially far--reaching, and
will be intimately connected with the processes occurring in the
innermost regions of the flow. As we have shown here, a crucial
ingredient in this respect is the angular momentum of the accreting
gas, even at values that are usually considered to be too low in terms
of the dynamics and the associated energy release.

\section{Discussion}\label{disc}

Spherical accretion flows on to black holes have in general low
radiative efficiencies, and this applies regardless of the cooling
mechanism. If the flow is slightly rotating, the situation changes
dramatically. Much of our effort in this work is therefore dedicated
to determining the state of the gas falling quasi--spherically onto
stellar--mass black holes with accretion rates in the range
$10^{-3}\lesssim \dot{M} \lesssim 0.5 M_{\sun}\;{\rm s}^{-1}$. In this
hypercritical regime, the gravitational accretion energy is carried
away by neutrinos \citep{chevalier89} .

Whether a disk forms or not depends on the precise value of the
angular momentum, which is difficult to calculate as it depends on the
physics of angular momentum transport inside the progenitor star. An
approximate condition for disk formation is that the specific angular
momentum of the infalling gas $l$ exceeds $r_{\rm g}c$. In most
previous discussions, it has been assumed that $l\gg r_{\rm g}c$ in
order for a black hole to switch to a luminous state. We argue here
that flows in hyperaccreting, stellar mass disks around black holes
are likely to transition to a highly radiative state when their
angular momentum is just above the threshold for disk formation $l
\sim 2 r_{\rm g}c$. In this regime, a {\it dwarf} disk forms in which
gas rapidly spirals into the black hole because of general
relativistic effects without any help from horizontal viscous
stresses. Such a disk is different from its standard high-$l$
counterpart as regards to its dynamics, energy dissipation and emitted
spectrum.

It is extremely unlikely that the progenitors of GRBs are just very
massive, single WR stars. Special circumstances are almost certainly
needed. The prompt formation of a black hole can provoke the failure a
core collapse supernova, if successfully shock breakout depends on
delayed neutrino heating from the proto--neutron star. So perhaps one
important distinction between a GRB and an ordinary supernova is
whether a black hole or a neutron star is formed in the aftermath.
However, not all black hole formation events can lead to a GRB: if the
minimum mass of a single star that leads to the formation of a black
hole is as low as $25 M_\sun$, this would overproduce GRBs by a large
factor \citep[see][]{izzard04,podsi04}.

The most widely discussed additional element for GRB production is
angular momentum: a rapidly rotating core is thought to be an
essential ingredient in the collapsar model \citep{woosley93,mw99}.
Massive stars are generally rapid rotators on the main
sequence. However, there are many well--established mechanisms, such
as mass loss and magnetic torques, by which they can lose a
substantial amount of angular momentum during their
evolution. Therefore, it is not at all clear whether the cores of
massive single stars will ever be rotating rapidly at the time of
explosion \citep[e.g.][]{heger05}. Relaxing the need for rapid
rotation may indeed increase the estimated rate for the single star
channel. The specific angular momentum $l$ of the accreting core is
thus a key factor. Evolutionary models \citep{heger05} indicate that
at $M=4 M_{\sun}$, the equatorial angular momentum of the
pre--supernova core inside a star with $M=15 M_{\sun}$ is $l_{0}
\approx 3 \times 10^{16} $~cm$^{2}$~s$^{-1}=0.8 r_{g} c$, if one
considers the effects of magnetic torques, and $l_{0} \approx 30
\times 10^{16} $~cm$^{2}$~s$^{-1}=8 r_{g} c$ if these are ignored. For
more massive stars the angular momentum is reduced even further (by
approximately one order of magnitude for a star with $M=25M_{\sun}$).
Clearly, relatively low values of angular momentum may be quite common
in the interior of these stars.

The simplest way to generate a high-$l$ disk ($l \gg r_{\rm g}c$) is
for the core to reside in a tight binary and be in corotation with the
binary.  The requirement that the total core angular momentum exceed
the maximum angular momentum of a Kerr black hole places interesting
limits on the binary period \citep{izzard04,podsi04}. If we make the
reasonable assumption that the binary is circular, and model the core
as an $n=3$ polytrope, the binary period must be smaller than $P_{\rm
orb} \sim 4 (M_{\rm core}/2 M_\sun)^{-1}(R_{\rm core}/10^{10}\;{\rm
cm})^{2}$ h. This orbit could be tight enough that the core may in
fact have been stripped of its helium in a common envelope to form a
CO core. By contrast, in most cases, the core of a massive, single
star is unlikely to retain the required angular momentum as its outer
hydrogen layers are blown off in a stellar wind. In this case, the
accretion flows would be quasi--spherical with $l\sim r_{\rm
g}c$. While these progenitor models were previously excluded on
grounds of low angular momentum, we suggest here that they should be
reconsidered as valid candidates. In fact, given the range of $l$
explored in this work, we argue that, as along as $l \geq 2r_{\rm
g}c$, low angular momentum cores may in fact be better suited for
producing neutrino-driven explosions following massive core collapse.

\acknowledgments We gratefully acknowledge helpful discussions with
A. Heger, T. Janka, A. MacFadyen, J. McKinney, P. M\'{e}sz\'{a}ros,
D. Proga, M. Rees and S. Woosley. We thank the referee for a prompt
and thorough report which helped prepare the final version. Financial
support for this work was provided in part by CONACyT--36632E (W.H.L.)
and NASA through a Chandra Postdoctoral Fellowship award PF3-40028
(E.R.-R.).


\clearpage

\clearpage

\clearpage

\clearpage

\clearpage

\clearpage

\clearpage

\clearpage

\clearpage

\clearpage

\begin{deluxetable}{ccccc}
\tablecolumns{4}
\tablewidth{0pc}
\tablecaption{Neutrino energies and maximum densities\tablenotemark{a}}
\tablehead{
\colhead{$l_{0}/(r_{g}c)$} & \colhead{$E_{\nu}^{\rm pair}$(MeV)} & 
\colhead{$E_{\nu}^{\rm ann.}$(MeV)} 
& \colhead{$\rho_{\rm max}$ (g~cm$^{-3}$)} 
& \colhead{$kT$(MeV)}}
\startdata
2.1 & 11 & 12 & $ 3 \times 10^{10}$ & 3 \\
2.2 & 11 & 12 & $ 3 \times 10^{10}$ & 3 \\
2.5 & 5  & 12 & $ 3 \times 10^{9} $ & 3 \\
3.0 & 4  & 12 & $ 2 \times 10^{9} $ & 3 \\
\enddata
\tablenotetext{a}{For all the runs shown here, $\alpha=0$ and
$\dot{M}=0.5$~M$_{\sun}$~s$^{-1}$.}
\end{deluxetable}

\end{document}